\documentstyle[prl,aps,floats]{revtex}
\input psfig
\begin{document}

\draft
\preprint{FNAL-Pub-97/001-E}
\title{
\begin{flushright}
FNAL-Pub-97/001-E
\end{flushright}
A Precision Measurement of Electroweak
Parameters in Neutrino-Nucleon Scattering}
\author{
        K.~S.~McFarland,$^3$ C.~G.~Arroyo,$^2$ B.~J.~King,$^2$ L.~de~Barbaro,$^5$
        P.~de~Barbaro,$^7$ A.~O.~Bazarko,$^2$ R.~H.~Bernstein,$^3$
        A.~Bodek,$^7$ T.~Bolton,$^4$ H.~Budd,$^7$ L.~Bugel,$^3$ J.~Conrad,$^2$
        R.~B.~Drucker,$^6$ D.~A.~Harris,$^7$ R.~A.~Johnson,$^1$
        J.~H.~Kim,$^2$ T.~Kinnel,$^8$ M.~J.~Lamm,$^3$
        W.~C.~Lefmann,$^2$ W.~Marsh,$^3$ 
        C.~McNulty,$^2$ S.~R.~Mishra,$^2$ D.~Naples,$^4$
        P.~Z.~Quintas,$^2$ A.~Romosan,$^2$ W.~K.~Sakumoto,$^7$ H. Schellman,$^5$
        F.~J.~Sciulli,$^2$ W.~G.~Seligman,$^2$ M.~H.~Shaevitz,$^2$
        W.~H.~Smith,$^8$ P.~Spentzouris,$^2$ E.~G.~Stern,$^2 $
        M.~Vakili,$^1$ U.~K.~Yang,$^7$ J.~Yu,$^3$ and G.~P.~Zeller$^5$
}
\address{ 
$^1$ University of Cincinnati, Cincinnati, OH 45221 \hfill
$^2$ Columbia University, New York, NY 10027 \\
$^3$ Fermi National Accelerator Laboratory, Batavia, IL 60510 \hfill
$^4$ Kansas State University, Manhattan, KS 66506 \\
$^5$ Northwestern University, Evanston, IL 60208 \hfill
$^6$ University of Oregon, Eugene, OR 97403 \\
$^7$ University of Rochester, Rochester, NY 14627 \hfill
$^8$ University of Wisconsin, Madison, WI 53706 \\
}
\date{\today}
\maketitle
\begin{abstract}
The CCFR collaboration reports a precise measurement of electroweak
parameters derived from the ratio of neutral-current to
charged-current cross-sections in neutrino-nucleon scattering at
the Fermilab Tevatron.  This
ratio of cross-sections measures the neutral current couplings to
quarks, which implies a determination of $\sin^2\theta_W^{\rm \scriptstyle (on-shell)}=0.2236\pm0.0028{\rm\textstyle
(expt.)}\pm0.0030{\rm\textstyle (model)}$ for $m_{\rm \scriptstyle top}=175$~GeV, $m_{\rm \scriptstyle Higgs}=150$~GeV.
This is equivalent to $M_W=80.35\pm0.21$~GeV.  The good agreement of
this measurement with Standard Model expectations implies the
exclusion of additional $\nu\nu qq$ contact interactions at $95\%$
confidence at a mass scale of 1-8~TeV, depending on the form of the
contact interaction.
\end{abstract}
\twocolumn

In the early 1980's, accurate measurements of neutrino neutral-current
scattering cross-sections provided key input to the Standard Model's
predictions of the $W$ and $Z$ boson masses.  Even with the production
of copious on-shell $W$ and $Z$ bosons at high luminosity
$p\overline{p}$ and $e^+e^-$ colliders, neutrino-nucleon ($\nu N$)
scattering still provides a measurement of electroweak parameters, in
particular $M_W/M_Z$, with comparable precision.  More importantly,
the high precision comparison among these distinct electroweak
processes differing in $q^2$ by more than two orders of magnitude
provides a critical test of the theory and the possibility to search
for non-Standard Model contributions with very high mass scales or low
probabilities\cite{LangLM,KevDonna}.  The measurement presented here
represents the most precise determination of $\sin^2\theta_W$ from $\nu N$
scattering to date and supersedes the previous result from CCFR
\cite{ArroyoKing} due to increased statistics and improved evaluation
of systematic errors.

The neutral-current (NC) and charged-current
(CC) $\nu N$ deep inelastic scattering
differential cross-sections on an isoscalar target of light quarks are
related by
\begin{eqnarray}
\frac{d}{dxdq^2}\left( \sigma_{\rm \scriptstyle NC}^{\nu,\overline{\nu}}\right) & = &
    \left( u_L^2+d_L^2\right) \frac{d}{dxdq^2}
      \left( \sigma_{\rm \scriptstyle CC}^{\nu,\overline{\nu}} \right) \nonumber \\
  && +
    \left( u_R^2+d_R^2\right) \frac{d}{dxdq^2}
    \left( \sigma_{\rm \scriptstyle CC}^{\overline{\nu},\nu} \right) \label{eqn:rnu}
\end{eqnarray}
where $u_{L,R}$ and $d_{L,R}$ are the left(L) and right(R)-handed
couplings of the $Z^0$ to up and down quarks, respectively.
Small corrections to Eqn.~\ref{eqn:rnu} arise from massive quark
production suppression, CKM matrix effects, higher-twist processes,
electromagnetic and electroweak radiative corrections, and from any isovector
component in the target, including heavy quark seas.  Within the
Standard Model, these left and right-handed couplings are given by
$I^{(3)}_{\rm \scriptstyle Weak}-Q_{\rm \scriptstyle EM}\sin^2\theta_W$ and $-Q_{\rm \scriptstyle EM}\sin^2\theta_W$,
respectively, allowing a measurement of $\sin^2\theta_W$ from ratios of NC to
CC, and $\nu$ to $\overline{\nu}$ CC cross-sections.  Furthermore, if the
expression for $\sigma^\nu_{\rm \scriptstyle NC}$ in Eqn.~\ref{eqn:rnu} is
used to extract $\sin^2\theta_W$, it is
almost equal to $\sin^2\theta_W$ in the ``on-shell'' renormalization scheme
($\sin^2\theta_W^{\rm \scriptstyle (on-shell)}\equiv 1-M_W^2/M_Z^2$ to all orders), independent of $m_{\rm \scriptstyle top}$ and
$m_{\rm \scriptstyle Higgs}$\cite{MarSir,Stuart}.  Therefore, the measurement of $\sin^2\theta_W$
from $\nu N$ scattering, combined with the precise measurements of
 $M_Z$ from LEP\cite{pdg}, implies a measurement of $M_W$.
In addition, direct extraction of $u_{L,R}$ and $d_{L,R}$ also allows a
search for a variety of non-Standard Model processes through
comparison of the measurements from $\nu N$ scattering with
Standard Model expectations.

The CCFR detector consists of an $18$~m long, $690$~ton neutrino target
calorimeter with a mean density of $4.2$~g/cm$^3$, followed by an iron
toroid spectrometer.  The target calorimeter consists of 168 iron
plates, $3$m~$\times$~$3$m~$\times$~$5.1$cm apiece.  The active elements
are liquid scintillation counters spaced every two plates ($10.2$~cm
of steel) and drift chambers spaced every four plates ($20.4$~cm of
steel).  There are a total of 84 scintillation counters and 42 drift
chambers in the target.  The neutrino target is approximately
isoscalar, with a $5.67\%$ neutron excess.  The toroid spectrometer is
not directly used in this analysis.

The Fermilab Tevatron Quadrupole Triplet neutrino beam is created by
decays of pions and kaons produced when $800$~GeV protons hit a BeO
target.  A wide band of secondary energies is accepted by downstream
focusing magnets.  The target is located about $1.4$~km upstream of
the neutrino detector. The target and focusing train are followed by a
$0.5$~km decay region.
The interactions of the beam are predominantly from muon neutrinos
($86\%$) and anti-neutrinos (12\%), but also include a small
fraction of electron neutrinos ($2.3$\%).
The mean energies of the
$\nu_\mu$, $\overline{\nu}_\mu$, and $\nu_e(\overline{\nu}_e)$ events are $165$, $135$ and
$160$~GeV, respectively.  The mean $q^2$ exchanged in the neutrino
interactions used in this analysis was $-35$~GeV$^2$.

\begin{figure}
\centerline{
\psfig{figure=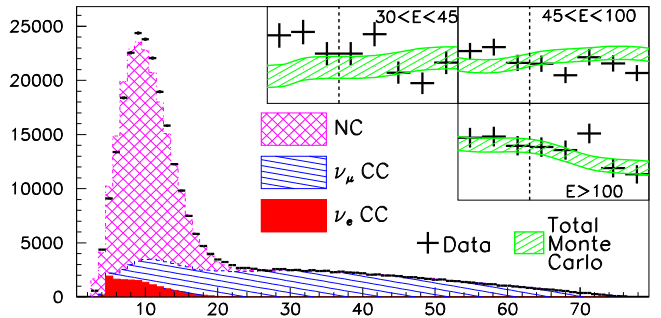,width=\columnwidth}}
\caption{Event length in data and Monte Carlo.  
The prediction for NC, CC $\nu_\mu(\overline{\nu}_\mu)$, and CC $\nu_e(\overline{\nu}_e)$
interactions is shown.  Inset
are comparisons in the region of the length cut 
of data and Monte Carlo with systematic errors shown.}
\label{fig:len}\typeout{need new length plots}
\end{figure}

Neutrinos are observed {\it via} their NC and CC interactions, both of
which are selected in this analysis from the energy transferred to the
struck quark which appears as a hadronic shower in the target
calorimeter.  The hadronic energy is measured by the variable $E_{\rm \scriptstyle cal}$,
which is the sum of energies observed in the first 20 scintillation
counters ($2.1$m equivalent of steel) in the target downstream of the
interaction vertex.  NC events usually have no final-state muon and
deposit energy over a range of counters typical of a hadronic shower
(5 to 20 counters).  $\nu_\mu$ CC events are distinguished by the
presence of a muon in the final state which deposits energy typical of
a minimum ionizing particle in a large number of consecutive
scintillation counters downstream of the hadronic shower.

A ``length'' is defined for each event as the number of
counters between the interaction vertex and the last counter
consistent with the energy deposition expected from a single muon
passing through the calorimeter.  Events with a ``short'' length are
identified as NC candidates.  The separation between short and
``long'' events is made at 20 counters ($2.1$m of steel) for events
with $E_{\rm \scriptstyle cal}\leq 45$~GeV, 25 counters ($2.6$m of steel) for events with
$45<E_{\rm \scriptstyle cal}\leq 100$~GeV, and 30 counters ($3.1$m of steel) for events
with $E_{\rm \scriptstyle cal}>100$~GeV.  As is shown in the length distributions in
Figure~\ref{fig:len}, NC interactions lie in a clear peak, well-below
the length cut, with a continuous band of CC $\nu_\mu$ interactions
under the peak.  The CC $\nu_\mu$ background is calculated to be
$10.5\%$, $21.3\%$, and $21.2\%$ of the ``short'' NC candidates in the
three $E_{\rm \scriptstyle cal}$ regions, respectively.

The data used in this analysis were taken between 1984 and 1988 in
FNAL experiments E-744 and E-770. Events were required to have
$E_{\rm \scriptstyle cal}>20$~GeV to ensure full efficiency of the
trigger\cite{ArroyoKing}.  Fiducial cuts were made on the location of
the neutrino interaction in the calorimeter to ensure that events were
neutrino-induced, that a separation between long and short events
could be made, and that events originated in the central part of the
calorimeter to maximize containment of wide-angle muons and to minimize
the ratio of electron to muon neutrinos.
The resulting data sample consisted of $8.1\times10^5$
events, and from these events the ratio
$$
R_{\rm \scriptstyle meas} = \frac{\rm\textstyle \#~of~short~events}{\rm\textstyle \#~of~long~events} = 0.4151\pm0.0010
$$
was calculated.

A detailed Monte Carlo was used to determine electroweak parameters
from $R_{\rm \scriptstyle meas}$.  The only undetermined inputs to this
Monte Carlo were the neutral current quark couplings which were then
varied until the Monte Carlo predicted an $R_{\rm \scriptstyle meas}$ which agreed with
that observed in the data.  For the extraction of $\sin^2\theta_W$, the
couplings in the Monte Carlo were fixed to their Standard Model
predictions as functions of $\sin^2\theta_W$, which was then varied as the only
free parameter.  The Monte Carlo included detector response and beam
simulations, as well as a detailed cross-section model which included
electromagnetic radiative corrections, isovector target corrections,
heavy quark production and seas, the longitudinal cross-section, and
lepton mass effects.

There are three major uncertainties in the comparison of $R_{\rm \scriptstyle meas}$ from
the Monte Carlo to the data: the statistical error in the data, the
uncertainty in the effective charm quark mass for charged current
charm production, and the uncertainty in the incident flux of $\nu_e$'s
on the detector.

The charm quark mass error comes from the uncertainty in modeling the
mass threshold of the charm production cross section.  The Monte Carlo
uses a slow-rescaling model with the parameters extracted using events
with two oppositely charged muons (e.g., $\nu q\to\mu^- c$, $c\to\mu^+
X$) from this experiment\cite{SAR}.  The shape and magnitude of the
strange sea were also extracted in the same analysis and were used in the
Monte Carlo cross-section model.  This error dominates the calculation
of $R_{\rm \scriptstyle meas}$ at low $E_\nu$ (and low $E_{\rm \scriptstyle cal}$) where the threshold
suppression is greatest.

The $\nu_e(\overline{\nu}_e)$ flux uncertainty has an important effect on $R_{\rm \scriptstyle meas}$
because almost all charged current $\nu_e(\overline{\nu}_e)$ events are short
events.  Therefore, the relatively small fractional uncertainty in the
$\nu_e(\overline{\nu}_e)$ flux is a significant effect, particularly at high $E_{\rm \scriptstyle cal}$
since most $\nu_e(\overline{\nu}_e)$ charged current interactions deposit the full
incident neutrino energy into the calorimeter.  Two techniques were
used to determine the $\nu_e(\overline{\nu}_e)$ flux.  In both E744 and E770, a
detailed beam Monte Carlo was used to predict the flux, up to a
$4.1\%$ uncertainty in each experiment.  This $4.1\%$ is
dominated by a $20\%$ production uncertainty in the $K_L$ content of
the secondary beam which produces $16\%$ of the $\nu_e$ flux.  The
bulk of the $\nu_e$ flux comes from $K^{\pm}_{e3}$ decays, which are
well-constrained by the observed $\nu_\mu$ spectrum from $K^{\pm}_{\mu
2}$ decays\cite{ArroyoKing}.  In E770, the $\nu_e(\overline{\nu}_e)$ flux was
measured directly using the fact that CC $\nu_e(\overline{\nu}_e)$ interactions
will have a high fraction of their energy deposited in the first three
counters downstream of the event vertex.  This gave an independent
measurement of the $\nu_e(\overline{\nu}_e)$ flux with a uncertainty of
$3.5\%$ which was in good agreement with the Monte Carlo
method\cite{Alex}.  Combining these techniques, a measurement of the
$\nu_e(\overline{\nu}_e)$ flux sum in E744 and E770 is obtained with a $2.9\%$
error.

\begin{table}[bt]
\begin{tabular}{|r|c|}
  SOURCE OF UNCERTAINTY & $\delta\sin^2\theta_W$ \\
\hline
 { data statistics } &  {0.0019}  \\
 { Monte Carlo statistics } & {0.0004} \\ \hline
 { TOTAL STATISTICS \hfill } & {0.0019} \\ \hline\hline
  { $\nu _e$ flux } &  {0.0015} \\
  { Transverse Vertex } & {0.0004} \\
  { Energy Measurement  \hfill } & \\
  { Hadron Energy Scale ($1\%$)} & {0.0004} \\
  { Muon Energy Loss in Shower} & {0.0003} \\
  { Muon Energy Scale ($1\%$)} & {0.0002} \\
  { Event Length     \hfill } & \\
  { Hadron Shower Length} & {0.0007} \\
  { Counter Efficiency and Noise} & {0.0006} \\
  { Vertex Determination} & {0.0003} \\ \hline
  { TOTAL EXP. SYST. \hfill } & {0.0019} \\ \hline\hline
  { Charm Production, $\overline{s}$} &   \\
  { ($m_c=1.31\pm0.24$~GeV)} & {0.0027}  \\
   { Higher Twist} & {0.0010} \\
  { Longitudinal Cross-Section} & {0.0008}  \\
  { Charm Sea, ($\pm100\%$)} & {0.0006} \\
   { Non-Isoscalar Target} & {0.0004} \\
   { Structure Functions} & {0.0002} \\ 
  { Rad. Corrections} & {0.0001} \\ \hline
  { TOTAL PHYSICS MODEL \hfill } & {0.0030} \\  \hline\hline  
  { TOTAL UNCERTAINTY \hfill } & {0.0041} \\
\end{tabular}
\caption{Uncertainties in the extraction of $\sin^2\theta_W^{\rm \scriptstyle (on-shell)}$ from
the CCFR data}\typeout{check table again}
\label{tab:uncer}
\end{table}

Other sources of experimental uncertainties were kept small through
extensive modeling based on neutrino and testbeam
data\cite{Sakumoto,King,Merritt}.  The cross-section model used a
modified Buras-Gaemers parameterization\cite{BGpar} of the CCFR data
for input parton distributions.  This resulted in partial
cancellations of certain systematic effects, such as errors in energy
calibration.  Systematic uncertainties associated with the measurement
of $E_{\rm \scriptstyle cal}$ include possible small NC/CC shower differences
(constrained by a LEPTO Monte Carlo study\cite{ArroyoKing}), 
uncertainties in the muon
energy deposit within the hadron shower, uncertainties in the
resolution function, $e/\pi$ response, and absolute energy scales
obtained from hadron and electron test beam
measurements\cite{Sakumoto,King}.  The length uncertainties include
those associated with the shower length parameterizations of test
beam measurements\cite{Merritt}, the calorimeter longitudinal vertex
determination (studied using the vertex from events with two muon tracks),
counter inefficiencies and noise, and counter spatial
dimensions.  
In the cross-section model, the level of the charm sea was taken from
the CTEQ4L parton distribution functions and was assigned a $100\%$
uncertainty.  Our parameterization of $R_{\rm \scriptstyle
long}=\sigma_L/\sigma_T$ is based on QCD predictions and data from
charged lepton scattering experiments\cite{Whitlow} and is varied by
$15\%$ of itself to estimate uncertainties.  A correction for the
difference between $u$ and $d$ valence quark distributions in
nucleons, obtained from muon scattering data\cite{NMC}, was applied to
account for the 5.67\% excess of neutrons over protons in the target.
A correction was also applied for the asymmetry in the $u$ and $d$ sea
distributions suggested by the NA51 Drell-Yan data\cite{DrellYan} and the
Gottfried Sum Rule as measured in muon scattering\cite{Gott}.
Radiative corrections to the scattering cross-sections were applied using
computer code supplied by Bardin\cite{Bardin} and
1-loop electroweak radiative corrections as calculated by Marciano and
Sirlin\cite{MarSir}.  Possible higher-twist
corrections were parameterized in a VDM-based model of
Pumplin\cite{Pumplin} which was constrained by lepto-production data.
Table~\ref{tab:uncer} shows the uncertainties in the determination of
$\sin^2\theta_W$.

\begin{figure}
\centerline{
\psfig{figure=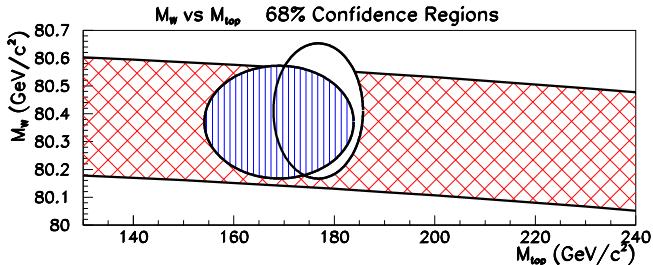,width=\columnwidth}}
\caption{$M_W$ vs $M_{\rm \scriptstyle top}$.  This measurement is shown as a cross-hatched band.  CDF
measurements provide the hollow ellipse, and the D0 measurements are shown in
the striped ellipse.}
\label{fig:mt-mw}
\end{figure}

The extraction of $\sin^2\theta_W^{\rm \scriptstyle (on-shell)}$ by comparing $R_{\rm \scriptstyle meas}$ in the data to the
Monte Carlo with $\sin^2\theta_W^{\rm \scriptstyle (on-shell)}$ as the single free parameter yields,
\begin{eqnarray}
\sin^2\theta_W^{\rm \scriptstyle (on-shell)}&=&0.2236\pm0.0028({\rm\textstyle(expt.)}\pm0.0030{\rm\textstyle (model)} \nonumber \\
&&+0.0006\times\left( \frac{(m_{\rm \scriptstyle top}^2-(175~{\rm\textstyle GeV})^2)}{(100~{\rm\textstyle GeV})^2}\right ) \nonumber \\
&& -0.0002\times\log_e\left( \frac{m_{\rm \scriptstyle Higgs}}{150~{\rm\textstyle GeV}}\right) .
\end{eqnarray}
The explicit dependence of the central value of this result on $m_c$
is $0.2236+0.0111\times(m_c-1.31~{\rm\textstyle GeV})$.  Only data with
$E_{\rm \scriptstyle cal}>30$~GeV was used in this result to reduce the effect of the
charm-production and higher-twist systematics which are largest at low
$E_{\rm \scriptstyle cal}$.  Since $\sin^2\theta_W^{\rm \scriptstyle (on-shell)}\equiv1-M_W^2/M_Z^2$, this result implies
$M_W=80.35\pm0.21$~GeV.  This value of $M_W$ agrees with direct mass
measurements as shown in Figure~\ref{fig:mt-mw}, and this result is
also in good agreement both with previous $\nu N$ measurements and
with Standard Model expectations\cite{pdg}.

\begin{figure}
\centerline{
\psfig{figure=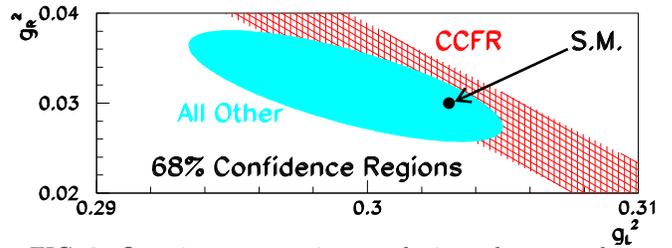,width=\columnwidth}}
\caption{One-sigma constraints on the isoscalar neutral current quark 
  couplings, $g_L^2$ and $g_R^2$, from this result (hatched) and from
other neutrino data (solid).}
\label{fig:coupl}\typeout{coupling figure old, need SM prediction}
\end{figure}

$R_{\rm \scriptstyle meas}$ is also used to
extract a constraint on the couplings of quarks to the $Z^0$:
\begin{eqnarray}
\kappa&=&0.5820\pm0.0041=1.7897g_L^2+1.1479g_R^2 \nonumber \\
&&-0.0916\delta_L^2-0.0782\delta_R^2,
\label{eqn:coupcon} 
\end{eqnarray}
where $g_{L,R}^2=u_{L,R}^2+d_{L,R}^2$ and
$\delta_{L,R}^2=u_{L,R}^2-d_{L,R}^2$.  The explicit dependence of the
central value of $\kappa$ on $m_c$ is
$\kappa=0.5820-(m_c-1.31)*0.0111$
The Standard Model prediction is $\kappa=0.5817\pm0.0013$ for the
measured values of $m_Z$, $m_{\rm \scriptstyle top}$, $m_W$.
Figure~\ref{fig:coupl} compares this result to
other neutrino data\cite{FogliHaidt} and the Standard Model
prediction\cite{pdg}.

\begin{table}[bt]
\begin{tabular}{|r|c|c|}
  Interaction & $\Lambda^+$ & $\Lambda^-$ \\
\hline
 LL & $4.7$~TeV & $5.1$~TeV \\
 LR & $4.2$~TeV & $4.4$~TeV \\
 RL & $1.3$~TeV & $1.8$~TeV \\
 RR & $3.9$~TeV & $5.2$~TeV \\
 VV & $8.0$~TeV & $8.3$~TeV \\
 AA & $3.7$~TeV & $5.9$~TeV \\
\end{tabular}
\caption{95\% Confidence Lower Limits on mass scales of new $\nu\nu qq$ contact terms from CCFR.}
\label{tab:limits}
\end{table}

Because the CCFR result for neutral-current quark couplings are in
good agreement with Standard Model expectations, this result disfavors
the introduction of additional processes contributing to the same
final state.  To quantify this, we use the CCFR data to constrain
models of new $\nu\nu qq$ contact interactions.  Such models are often
used to parameterize searches for fermion compositeness or non-zero
size\cite{EichtenLanePeskin}.  We assume a generation-universal interaction of the form
$$
-{\cal L}=\sum_{H\in\{L,R\}} \frac{\pm 4\pi}{\left( \Lambda^\pm_{H_lH_q}\right) ^2}
\overline{l}_{H_l}\gamma^\mu l_{H_l}\overline{q}_{H_q}\gamma_\mu q_{H_q}.
$$
Such interactions would shift one or more of the quark couplings in
$\nu$ or $\overline{\nu}$ induced interactions from its Standard Model 
value\cite{LangLM}.
One-sided $95\%$ confidence level lower limits for each $\Lambda$ are
set by finding the points in the space of measured couplings at which
the $\chi^2$ is $1.64$ units above the minimum, and then determining
the $\Lambda$ to which these points correspond in each model.  Limits
for LL, LR, RL, RR, VV (vector,
$\Lambda_{LL}=\Lambda_{LR}=\Lambda_{RL}=\Lambda_{RR}$) and AA
(axial-vector,
$\Lambda_{LL}=-\Lambda_{LR}=-\Lambda_{RL}=\Lambda_{RR}$) are shown in
Table~\ref{tab:limits}.  These limits are roughly comparable to the
limits for charged lepton-quark interactions in
$p\overline{p}$ collider data from the Fermilab Tevatron\cite{CDF}.

In summary, CCFR has produced the most precise measurements of
neutral-current neutrino-nucleon interactions to date, and has used these
measurements to constrain neutral-current coupling to quarks.  Within
the Standard Model, this leads to a measurement of
$\sin^2\theta_W^{\rm \scriptstyle (on-shell)}=0.2236\pm0.0041$ ($m_{\rm \scriptstyle top}=175$~GeV, $m_{\rm \scriptstyle Higgs}=150$~GeV) which
corresponds to $M_W=80.35\pm0.21$~GeV.  This result
is also used to limit possible TeV-scale contact interactions of
neutrinos and quarks outside the Standard Model.

\end{document}